\documentclass[12pt]{article}
\usepackage{epsfig}
 
\newlength{\abstwidth}
\begin{document}
 
\sloppy
 
\pagestyle{empty}
 
\begin{flushright}
LU TP 97--18 \\
August 1997
\end{flushright}
 
\vspace{\fill}
 
\begin{center}
{\LARGE\bf A matching of}\\[3mm]
{\LARGE\bf matrix elements and parton showers}\\[10mm]
{\Large J. Andr\'e and %
T. Sj\"ostrand\footnote{torbjorn@thep.lu.se}} \\[3mm]
{\it Department of Theoretical Physics,}\\[1mm]
{\it Lund University, Lund, Sweden}
\end{center}
 
\vspace{\fill}
 
\begin{center}
{\bf Abstract}\\[2ex]
\begin{minipage}{\abstwidth}
We propose a simple scheme to start a parton-shower evolution 
description from a given jet configuration in $e^+ e^-$ annihilation
events. This allows a convenient combination of the full angular
information content of matrix elements with the detailed sub-jet 
structure of parton showers, and should give a realistic overall 
description of event properties. Explicit studies with this
hybrid approach are presented for the four-jet case, as a simple 
testing ground of the ideas. 
\end{minipage}
\end{center}
 
\vspace{\fill}
 
\clearpage
\pagestyle{plain}
\setcounter{page}{1}
In modern high-energy particle physics, the ability to perform
precision measurements or tests of fundamental concepts often 
requires an accurate modelling of multiparticle topologies.
These events have their origin in the perturbative production of 
sets of partons --- quarks and gluons --- that subsequently hadronize
nonperturbatively to produce the observable final state. The
first step therefore is to obtain an accurate description of
the perturbative production of multiparton topologies. 
Basically, there exists two approaches to this problem.

The standard method is the \emph{matrix element} one.
In this approach the relevant Feynman graphs are calculated,
order by order in $\alpha_{\mathrm{s}}$, until the required 
accuracy is ensured. Taking the convenient example of the 
lowest-order process $e^+ e^- \to \gamma/Z^0 \to q\overline{q}$,
the first-order real production process is  
$e^+ e^- \to \gamma/Z^0 \to q \overline{q} g$. The infrared 
and collinear gluon-emission divergences of the latter process
are cancelled by the first-order virtual corrections to the
lowest-order graph, to give a finite total cross section.
In second order, two new processes appear, 
$e^+ e^- \to \gamma/Z^0 \to q \overline{q} g g$ and
$e^+ e^- \to \gamma/Z^0 \to q \overline{q} q \overline{q}$,
together with virtual corrections to the lower-order processes.

In principle, the matrix-element approach is the correct one,
that gives the full information content of perturbative QCD.
There are caveats, however. One is that the calculations
become increasingly complex in higher orders. This in particular
affects the virtual corrections, where subtle cancellations of
singularities have to be performed between graphs corresponding
to different number of phase-space dimensions. Therefore a full 
calculation of all virtual corrections only exists to second order 
in $\alpha_{\mathrm{s}}$ \cite{ert}.
Another caveat is that the matrix-element method is unreliable
when applied to the exclusive production of two nearby partons. In an
order-by-order calculation of such a fixed parton configuration,
the virtual corrections are (to leading-log accuracy) expected to
appear as an alternating series that sums up to give an exponential
damping --- a Sudakov form factor \cite{sudakov} . The closer a 
gluon is to another parton, the larger are the terms of this series,
and the more terms are necessary to obtain a convergent answer.

The alternative approach is the \emph{parton shower} one. 
Here the amplitude-based language of matrix elements is replaced 
by a simplified probabilistic one. The basic process
$e^+ e^- \to \gamma/Z^0 \to q \overline{q}$ is supplemented by 
standard QCD evolution-equation rules for subsequent branchings 
$q \to q g$, $g \to g g$ and $g \to q \overline{q}$. By repeated 
application of these simpler $1 \to 2$ branchings arbitrarily 
complicated multiparton topologies are generated. The shower 
approach obeys detailed balance --- a branching transforms an 
$n$-parton configuration to an $n+1$ one --- which implies that 
a Sudakov form factor enters in a natural fashion. While the
traditional parton-shower approach formally is of leading-log
accuracy only, in reality many next-to-leading effects are 
included: an event generator takes full account of energy--momentum
conservation and recoil effects, it can be made to include
angular ordering (coherence) of emissions \cite{coher}, the 
$\alpha_{\mathrm{s}}$ scale can be optimized based on knowledge
of higher-order branching kernels \cite{alphas}, and azimuthal angles 
in branchings can be chosen non-isotropically to include both spin 
and coherence effects \cite{azimuth}.

However sophisticated, the parton-shower approach still cannot be
expected to cover the full information content available in the
matrix-element expressions. This should be especially notable    
for situations when partons are well separated: here exact 
kinematics is important and several graphs can be expected 
to contribute comparably much, so that interference terms are 
significant. 

The picture then is one where matrix elements are likely to provide
the better description of the main character of events, i.e. the
topology of well separated jets, while parton showers should be
better at describing the internal structure of these jets. A
marriage of the two approaches seems rather natural, but is 
actually not so simple. Only for first-order $e^+ e^-$ events is 
this routinely done. One approach \cite{mbts} is to generate 
a normal parton shower starting from a $q \overline{q}$ 
topology, and reweight (by a rejection technique) the first branchings 
of the $q$ and the $\overline{q}$ so that the first-order 
$q \overline{q} g$ three-jet matrix elements are reproduced.
An improvement is to include weights in all branches \cite{mike}. 
A clever choice of cascading description can automatically 
lead to a good agreement also with four-jet matrix elements 
\cite{ariadne}. Another technique is to use matrix elements
to generate a varying number of initial partons that then are 
allowed to shower further \cite{herwig}. This strategy has also 
been applied to deep inelastic scattering \cite{lepto}.
The number of event classes to consider can be reduced if
cut-offs are chosen appropriately \cite{reno}.

In this letter we propose another strategy for combining matrix 
elements with parton showers. It has the advantage that it can be 
applied to arbitrarily complicated partonic states, but the 
disadvantage that it does not tell how to mix different event topologies 
consistently. Its main application therefore is to events where 
the bulk properties are given by matrix elements, i.e. where the main 
partons are well separated, and the task is to provide a realistic 
representation of the structure of the resulting jets. One such 
example is four-jet studies at LEP~1 to test the coupling structure 
of QCD, another four-jet studies at LEP~2 to control backgrounds to 
$W^+ W^-$ events.  

The basic idea is to cast the output of matrix element generators in 
the form of a parton-shower history, that then can be used as input 
for a complete parton shower. In the shower, that normally would be 
allowed to develop at random, some branchings are now fixed to
their matrix-element values while the others are still allowed
to evolve in the normal shower fashion. The preceding history of the 
event is also in these random branchings then reflected e.g. in terms 
of kinematical or dynamical (e.g. angular ordering) constraints.  
 
\begin{figure}
\begin{center}
\mbox{\epsfig{file=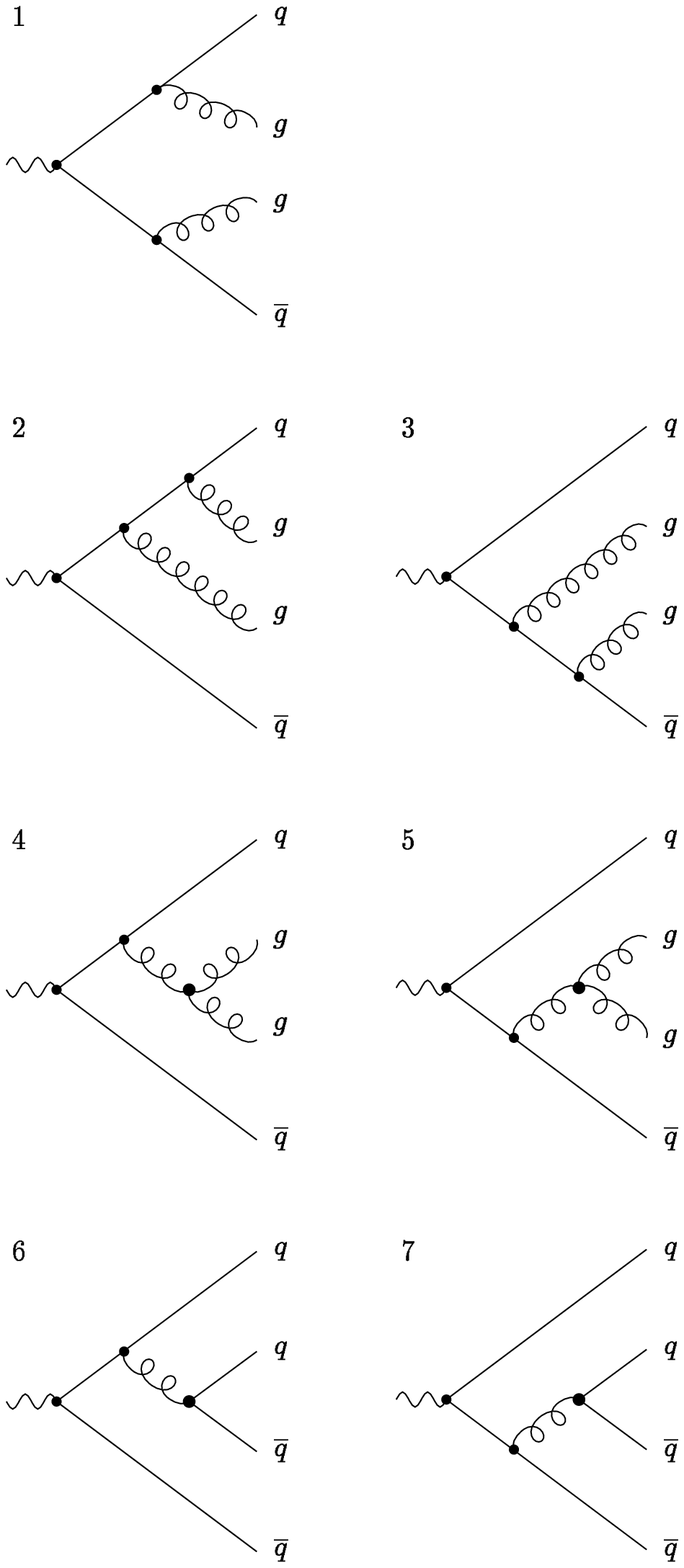,height=20cm,bbllx=0,bblly=140,%
bburx=400,bbury=810}}
\end{center}
\caption[]{4-parton histories to second order in $\alpha_{\mathrm{s}}$}
\label{fig1}
\end{figure}

\begin{figure}
\begin{center}
\mbox{\epsfig{file=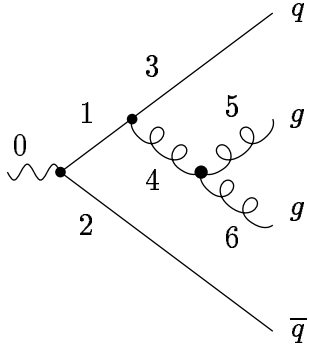,bbllx=0,bblly=630,%
bburx=300,bbury=790}}
\end{center}
\caption[]{One 4-parton history}
\label{fig2}  
\end{figure}

The idea is best exemplified by $e^+ e^-$ four-jet events. The process 
receives contributions from seven graphs, see Fig. \ref{fig1}. The
$q \overline{q} g g$ and $q \overline{q} q \overline{q}$ graphs easily 
separate, so we can concentate on the former processes. Then the 
matrix-element expression contains contributions from five graphs, 
and from interferences between them. The five 
graphs can also be read as five possible parton-shower histories for 
arriving at the same four-parton state, but here without the 
possibility of including interferences. The relative probability for 
each of these possible shower histories can be obtained from the rules 
of shower branchings.
For example, the relative probability for the history shown in
Fig.~\ref{fig2} is given by:
\begin{equation}
{\cal P} = {\cal P}_{1\rightarrow 34} {\cal P}_{4\rightarrow 56}
= \frac{1}{m_{1}^{2}} \frac{4}{3} \frac{1+z^{2}_{34}}{1-z_{34}} 
\cdot \frac{1}{m_{4}^{2}} 3 
\frac{(1-z_{56}(1-z_{56}))^{2}}{z_{56}(1-z_{56})}
\end{equation}
where the probability for each branching contains the mass 
singularity, the colour factor and the momentum splitting
kernel. The masses are given by
\begin{eqnarray}
m_{1}^{2} = p_{1}^{2} & = & (p_{3}+p_{5}+p_{6})^{2}~, \\
m_{4}^{2} = p_{4}^{2} & = & (p_{5}+p_{6})^{2}~, \nonumber
\end{eqnarray}
and the z values by
\begin{eqnarray} 
z_{bc} = z_{a \to bc} &=& 
\frac{ m^{2}_{a} }{ \lambda }\frac{ E_{b} }{ E_{a} } - 
\frac{m^{2}_{a} - \lambda + m^{2}_{b} - m^{2}_{c}}{2\lambda}
\\
\mathrm{with~~}
\lambda &=& \sqrt{(m^{2}_{a} - m^{2}_{b} - m^{2}_{c})^{2} - 4m^{2}_{b}\,
m^{2}_{c}}~.
\nonumber
\end{eqnarray}
The form of the probability matches the expression used in the
{\sc Jetset} parton-shower algorithm \cite{jetset}.
Other programs use the opening angle \cite{herwig} or 
the transverse momentum \cite{ariadne} as main evolution variable
instead of mass. The $z$ definition (again the one used in {\sc Jetset}) 
reduces to energy fractions in the limit that the daughters are massless, 
and corresponds to unchanged decay angle in the rest frame of the mother 
when daughter masses are introduced. 

Variants on the above probabilities are imaginable. For instance, in the
spirit of the matrix-element approach we have assumed a common 
$\alpha_{\mathrm{s}}$ for all graphs, which thus need not be shown, 
whereas the parton-shower language normally assumes $\alpha_{\mathrm{s}}$
to be a function of the transverse momentum of each branching
\cite{alphas}. One could also include information on azimuthal 
anisotropies \cite{azimuth}.

The relative probability ${\cal P}$ for each of the five possible 
parton-shower histories can be used to select one of the possibilities 
at random. (A less appealing alternative would be a ``winner takes 
all'' strategy, i.e. selecting the configuration with the largest 
${\cal P}$.) The selection fixes the values of the $m$, $z$ and 
$\varphi$ at two vertices. The azimuthal angle $\varphi$ is defined 
by the daughter
parton orientation around the mother direction. When the conventional 
parton-shower algorithm is executed, these values are then forced on the
otherwise random evolution. This forcing cannot be exact for the $z$ 
values, since the final partons given by the matrix elements are on the 
mass shell, while the corresponding partons in the parton shower might 
be virtual and branch further. The shift between the 
wanted and the obtained $z$ values are rather small, very seldom
above $10^{-6}$. More significant are the changes of the opening
angle between two daughters: when daughters originally assumed 
massless are given a mass the angle between them tends to be reduced.
This shift has a non-negligible tail even above 0.1 radians. The 
``narrowing'' of jets by this mechanism is compensated by the 
broadening caused by the decay of the massive daughters, and thus
overall effects are not so dramatic.

All other branchings of the parton shower are selected at random 
according to the standard evolution scheme. In Fig.~2, this means 
that partons 2, 3, 5 and 6 (and any daughters) have random masses 
and branchings. There is an upper
limit on the non-forced masses from internal logic, however.
For instance, for four-parton matrix elements, the singular
regions are typically avoided with a cut $y > 0.01$, where $y$ is
the square of the minimal scaled invariant mass between any pair of 
partons. Larger $y$ values could be used for some purposes, while
smaller ones give so large four-jet rates that the need to
include Sudakov form factors can no longer be neglected. 
The $y > 0.01$ cut roughly corresponds to $m > 9$~GeV at LEP~1
energies, so the hybrid approach must allow branchings at least
below 9~GeV in order to account for the emission missing from the
matrix-element part. Since no 5-parton emission is generated by the 
second-order matrix elements, one could also allow a threshold
higher than 9 GeV in order to account for this potential emission.
However, if any such mass is larger than one of the forced masses, 
the result would be a different history than the chosen one.
Thus one plausible strategy is to choose as threshold the smallest 
of the two forced masses. Another choice is a fixed mass threshold
at 13 GeV, a value that gives the same average multiplicity as the
original parton shower. This may be viewed as a pragmatical compromise
between the two extremes above.

\begin{figure}
\begin{center}
\mbox{\epsfig{file=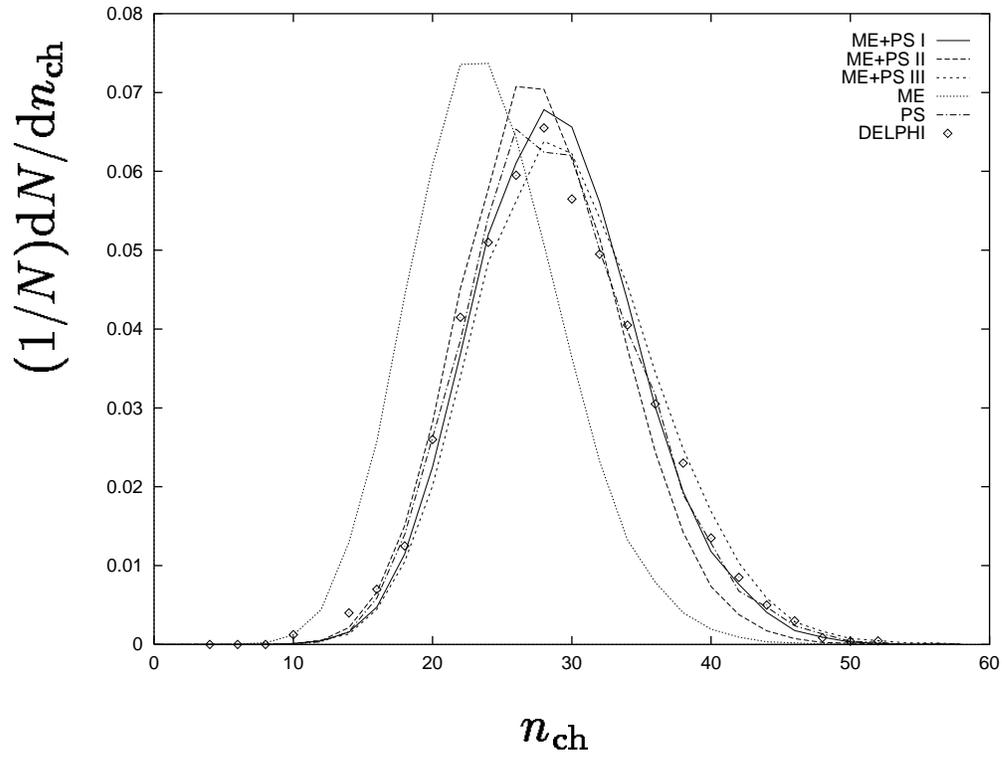,bbllx=50,bblly=500,%
bburx=500,bbury=800}}
\end{center}
\caption[]{Distributions of charged particle multiplicity}
\label{fig3}
\end{figure}

The resulting charged multiplicity distributions are shown in 
Fig.~\ref{fig3}. 
The JADE P0 scheme \cite{jade} has been used to find four-jet 
events with a separation $y > 0.02$, and results are compared with
DELPHI data \cite{delphi}. The hybrid is executed with three different 
mass thresholds: fixed at 13 GeV (ME+PS I), fixed at 9 GeV (ME+PS II) 
and minimum of the two forced masses (ME+PS III). In all cases the 
partons have been hadronized with the string fragmentation scheme
\cite{agis}. The distributions of the parton shower (PS) and the 
three hybrids nearly coincide, and agree with the data. 
The matrix elements distribution (ME) is peaked
at significantly lower multiplicities, as should be expected from
the lesser activity. This difference could be reduced by 
a retuning of the fragmentation parameters for the matrix-element
approach, but this retuning would then have to be done separately 
at each energy, while the parton-shower tuned parameters should be
valid at all energies and for all jet multiplicities.

\begin{figure}
\begin{center}
\mbox{\epsfig{file=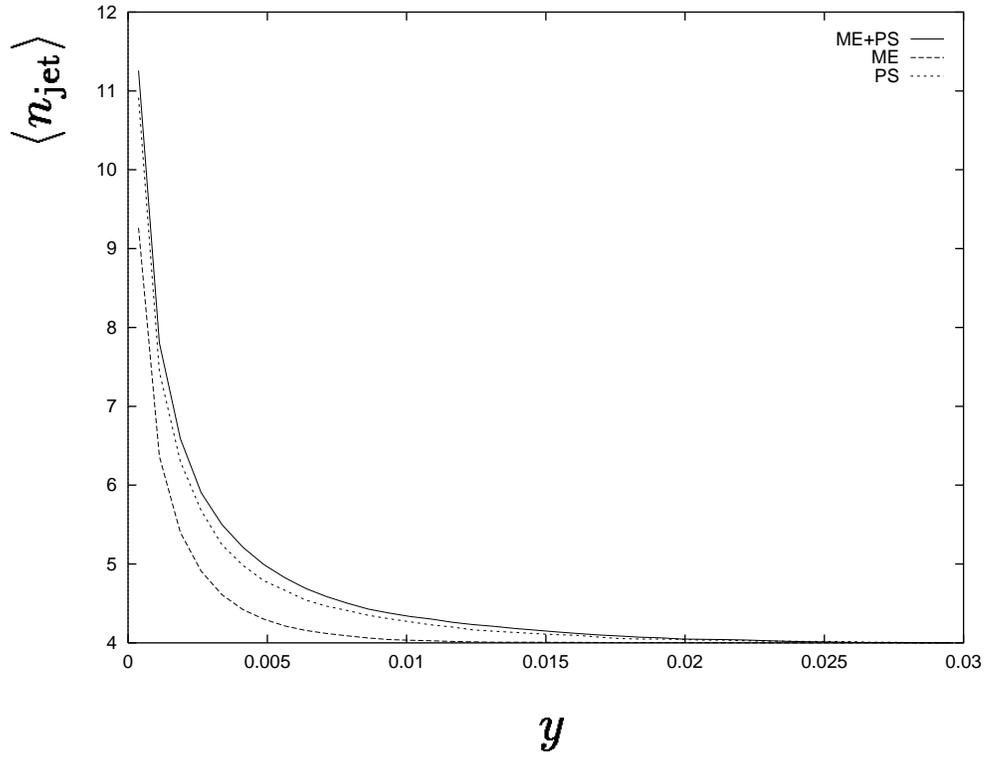,bbllx=50,bblly=500,%
bburx=500,bbury=800}}
\end{center}
\caption[]{Subjet multiplicities}
\label{fig4}
\end{figure}

The subjet multiplicity is shown in Fig.~\ref{fig4}. This is the
average number of jets as a function of $y < 0.03$, given that there
are exactly four jets at $y = 0.03$. Thus the subjet multiplicity 
gives a picture of the width of the four original jets. The ME+PS 
events have been generated using the fixed mass threshold at 13 GeV.
The curve corresponding to the matrix elements has the strongest tendency 
towards 4 jets, as could be expected, whereas ME+PS agrees rather well
with the convenventional parton shower.

\begin{figure}
\begin{center}
\mbox{\epsfig{file=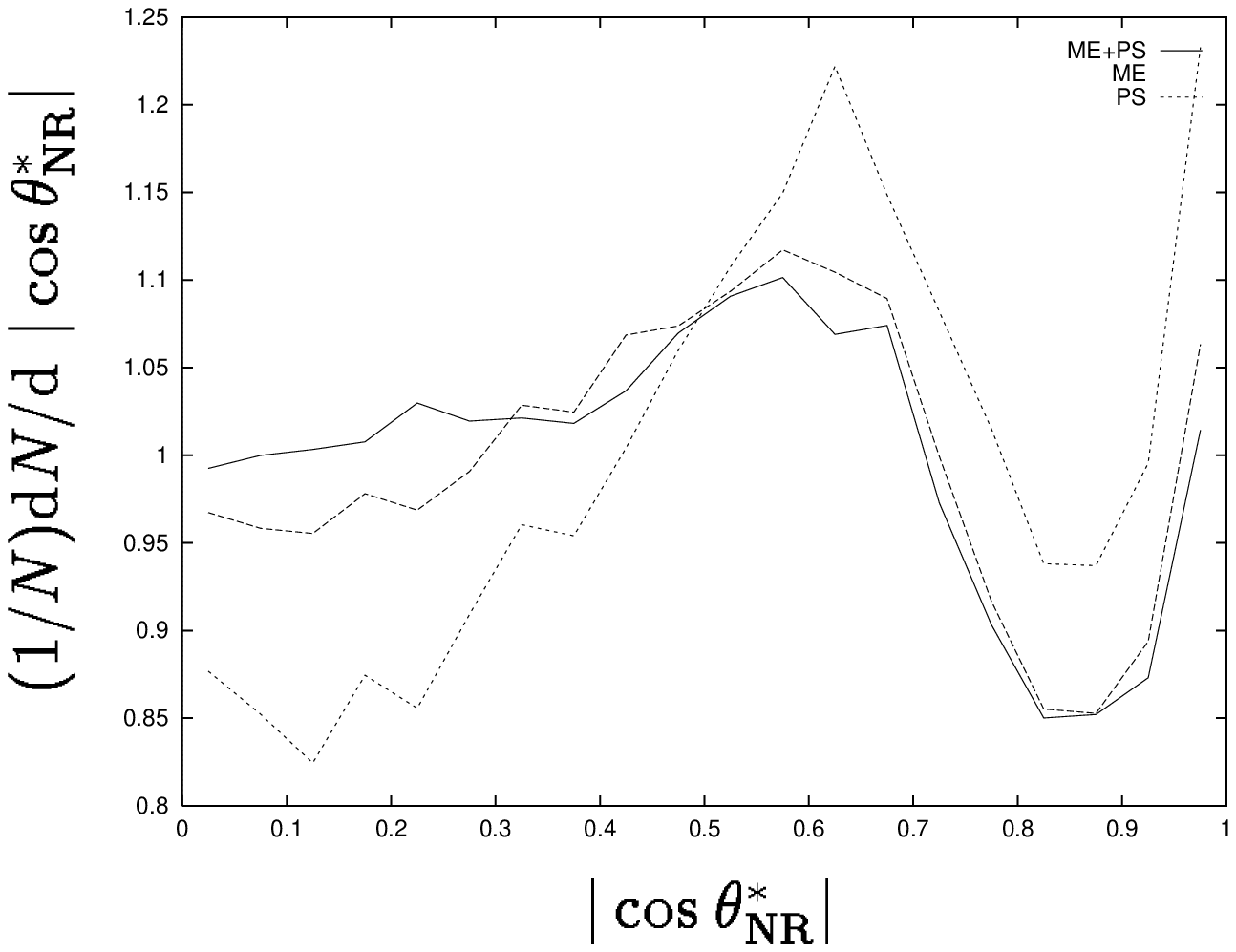,bbllx=50,bblly=500,%
bburx=500,bbury=800}}
\end{center}
\caption[]{$|\mathrm{\cos}\theta^{\ast}_{\mathrm{NR}}|$ distributions}
\label{fig5}
\end{figure}

Angular distributions in four-jet events have been studied in order
to test the predictions of QCD \cite{angles}. We therefore already know
that these distributions are well described by the matrix-element
approach, while the {\sc Jetset} parton shower is somewhat less
accurate \cite{angledata}. Without entering into the (not easily
reproducible) details of the experimental procedure, we can thus 
directly compare the hybrid approach with the ME and PS options. 
Fig.~\ref{fig5} gives the distribution of one such angle for 
four-jets at $y=0.03$. As expected, the ME+PS curve here agrees 
well with the pure ME one.

In summary, we have therefore obtained a simple scheme that allows a 
convenient combination of the best aspects of the matrix-element and 
parton-shower approaches to the description of multihadronic $e^+e^-$ 
annihilation events. The structure of well separated jets agrees with 
the  matrix-element prediction, while the richer substructure of jets 
follows the parton-shower pattern. Further details on the algorithm 
and tests of it can be found elsewhere\cite{johan}; the four-parton
program itself is also available \cite{program}.

The method has here been illustrated for four-jet events, but could
easily be extended to any other fixed multiplicity.
A possible weakness of the approach, however, is that it does not 
automatically tell how to mix different jet multiplicities in the 
proper proportions. One could use the matrix-element mixture as a 
starting point, and generate three- and two-jet events with a veto 
against too large masses, similarly to what was discussed above.
This would not be unreasonable, but the Sudakov form factors of the
pure parton-shower approach are then missing and therefore the
transition between jet multiplicities is not as well controlled.

The same kind of approach could also be applied to other processes,
such as three-jet production in hadron collisions, which could be
viewed as a basic $2 \to 2$ scattering combined with one forced
parton-shower branching. This branching could here occur either in 
the initial or in the final state, however, so there are more
possibilities to keep track of. Furthermore, the cross section of 
the $2 \to 2$ scattering here needs to be included in the relative 
probability to select a given configuration, since this scattering 
varies between the possible event histories.

\newpage

\end{document}